\documentstyle[aasms4,epsfig,psfig]{article}  
\def\simlt{\lower.5ex\hbox{\ltsima}}   
\def\gtsima{$\; \buildrel > \over \sim \;$}
\def\simgt{\lower.5ex\hbox{\gtsima}}

\begin{document}

\title{Forming giant planets via fragmentation of protoplanetary disks}

\author{Lucio Mayer $^1$, Thomas Quinn $^1$, James Wadsley$^2$, Joachim Stadel$^3$}
$^1$  Department of Astronomy, University of Washington,
Seattle, WA 98195, USA mayer@astro.washington.edu, trq@astro.washington.edu\\
$^2$ Department of Physics \& Astronomy, McMaster University, 1280 Main St. West, Hamilton ON L8S 4M1 Canada, wadsley@physics.mcmaster.ca \\
$^3$ University of Victoria, Department of Physics and Astronomy, 3800 Finnerty
Road, Elliot Building, Victoria, BC V8W 3PG Canada,stadel@phys.uvic.ca\\

\begin{abstract}

{\bf  The evolution of gravitationally unstable protoplanetary gaseous 
disks has been studied using three dimensional smoothed particle 
hydrodynamics (SPH) simulations with unprecedented resolution.
We have considered disks with initial masses and temperature profiles 
consistent with those inferred for the protosolar nebula and for other 
protoplanetary disks. We show that long-lasting,
self-gravitating protoplanets arise 
after a few disk orbital times if cooling is efficient enough to maintain the
temperature close to 50 K. The resulting bodies have masses and 
orbital eccentricities remarkably similar to those of observed extrasolar 
planets.}

\end{abstract}

\section{}

About 100 extrasolar planets have been detected by the wobble they induce 
on their star$({\it 1,2})$. Their masses range from about one Jupiter
mass ($M_J$) to more than 10 $M_J$ and have orbits ranging from 
nearly circular to very eccentric.  In the standard core-accretion
model giant planets might require longer than $10^6$ years to
form$({\it 3,4})$, which could exceed observed disk
lifetimes $({\it 5-7})$. In particular, more than 80\% of the stars in the 
Galaxy probably formed in
dense clusters like those in the Orion nebula$({\it 8})$ where the ultraviolet
radiation of bright stars can ablate the gaseous disk in far 
less than a million years$({\it 5,6})$. Hence giant planet formation
must occur quickly  or such planets would be rare.
Even in the case where a large solid core is assembled rapidly enough, 
torques acting between the disk and the protoplanets are believed to 
induce its complete inward migration in a few thousand years$({\it 9,10})$
 --- planets could sink towards the star before being able to accrete 
the large gaseous masses observed $({\it 11,12})$.
Alternatively, giant planets could coagulate directly in the gas
component as a result of gravitational instabilities in a cold disk with a 
mass comparable to that adopted in the core-accretion model$({\it 13,14})$.
Simulations done with codes that solve the hydrodynamical equations on 
a fixed grid show
that slightly perturbed disks form strong spiral arms and overdensities 
at $R > 10$ AU$({\it 15,16})$ where the temperature can be lower than 
60 K$({\it 17,18})$.  
The trigger of the instability 
might come from material of the protostellar cloud infalling onto the 
disk$({\it 13})$. 
If these condensations are long-lasting and 
can contract to planetary densities, gravitational instability
would be the prevailing  formation mechanism for giant planets
because it takes less than a thousand years$({\it 13,15})$.
Solid cores with masses as low as currently estimated
for Jupiter (between 0 and 10 Earth masses$({\it 19})$) 
could then form inside the
gaseous protoplanets due to dust and planetesimals driven there by
local pressure gradients in a few thousand years$({\it 20})$.

However, due to the limitations of the techniques, simulations  
have not yet been able to show convincingly that the overdensities are 
not sheared apart by the tidal field of the star, nor that they can collapse into protoplanets$({\it 16)}$. One needs to achieve a
high spatial resolution for such a purpose. 
Smoothed particle hydrodynamics (SPH) simulations$({\it 21})$
describe the gaseous medium as a collection of particles and can
follow arbitrarily high densities. Disks have been simulated
with this technique in the past but with less than $10^5$ particles
$({\it 22-24})$; with such
a low mass resolution the evolution of the density distribution is considerably noisy and artificial fragmentation can take place$({\it 23})$.

Here we report on the results of new 3D SPH simulations of marginally 
unstable disks of molecular hydrogen using as many as 1 million particles. 
The disks
extend from 4 to 20 AU initially and are in nearly keplerian rotation around a solar mass 
star represented by a point mass.
They have a minimum Toomre Q parameter of either $1.4$ or $1.75$, and masses 
of, respectively, $0.1$ and $0.08 M_{\odot}$ (see caption of Fig.1 for
details).
In the initial stage the disks are evolved using a locally isothermal 
equation of 
state; the initial temperature decreases with radius following a power 
law profile
predicted by detailed calculations of thermal balance between the central
star, the disk and the protostellar cloud$({\it 17})$ and is then held fixed
locally. This approximation is based on the assumption that the cooling time
is so short that the disk radiates away any thermal energy injection
on a time-scale shorter than the orbital time.
This seems to be supported by recent grid-based simulations that 
include radiative transfer in the diffusion approximation$({\it 16})$
but neglect the irreversible heating that can be generated by
shock waves in a strongly unstable disk$({\it 25,26})$.
While further investigation on the balance between heating and cooling
in realistic disks will be needed in the future $({\it 26})$, 
here we concentrate on showing that actual 
protoplanets can form if the disk remains cold long enough$({\it 27})$.

After $\sim 150$ years, corresponding to about 5 orbital times at a radius
of 10 AU, the disks develop trailing spiral arms and local
overdensities at $R > 10$ AU.
In the lighter disk the spiral arms grow in amplitude up to about 300 years, 
and then they settle down to a nearly stationary pattern (Fig.1).
In the most massive disk ($Q=1.4$) a two-armed mode grows in amplitude 
up to the point where, after about 200 years, fragmentation occurs along 
the arms, and more than one distinct clump appears (Fig.1); then 
additional strong arms appear even at $R \simgt 7-8$ AU, and more
clumps are formed. Inside this radius the disk is too hot for 
condensations to form.
Clumps
quickly contract, reaching central densities of more than $10^5$ times 
the local density in a matter of a few orbital times (tens of years).
These condensations are self-gravitating, their masses being larger than
the local Jeans mass $({\it 28})$, and easily resist stellar tides.

Such dense objects would be optically thick and 
would be unable to cool radiatively as efficiently as
assumed by the locally isothermal approximation$({\it 16,25})$.
Therefore, we run again the same initial conditions changing 
the equation of state to adiabatic$({\it 29})$ as soon as 
the spiral modes approach fragmentation,
namely when they reach a density around ten times higher than the initial 
local density$({\it 16,26})$. 
The equation of state is changed throughout the disk. 
After 350 years non-axisymmetric features in the disk are weaker
compared to the isothermal simulation, yet clump formation 
has proceeded.
Nearly as many clumps as in the locally isothermal run
are still present at $R > 10$ AU and have central densities 
still $10^5$ times higher than
the local density. The clumps are rapidly rotating spheroids; 
assuming conservation of angular momentum, bodies of about a Jupiter mass
would have a rotation period of a few hours if they were allowed to contract
further and reach the density of Jupiter.

The masses of the clumps shortly after all of them are in
place (after $\sim 350$ years) 
range from 1 to 5 $M_J$ (all gas at 
densities at least ten times higher than the initial 
local density is identified as a clump), comparable
to the masses of extrasolar planets$({\it 1})$.
Clumps formed along the same spiral arm collide and merge into a more 
massive object sooner or later.
In addition, all clumps accrete gas from the disk and begin to clear gaps.

Time-steps can be as small as a few hours inside the clumps, and this slows 
down the simulation considerably.
To keep following the clumps on a longer timescale we 
resort to simulations with a resolution 5
times lower. By 350 years 7 clumps with masses comparable to the largest
among the 13 clumps present in the higher 
resolution simulation have formed and are then followed for about 1000 years.
We run both a locally isothermal and an adiabatic simulation to determine
the role of the thermal structure of the disk in shaping the dynamics
of clumps. 

Clumps are born on orbits with a wide range of eccentricities driven by
the underlying strongly non-axisymmetric disk potential.
These orbits then evolve in a variety of ways (Fig.2).
The smoother disk in the adiabatic run leads to less eccentric orbits;
also, the higher pressure
of the gas reduces considerably gas accretion by clumps.
Slightly more circular orbits reduce the rate of close encounters and 
mergers between clumps relative to the isothermal run.
After nearly a thousand years three clumps
are left in the adiabatic run (Fig.2) as opposed to two clumps in the 
isothermal run. These numbers are comparable to those of 
extrasolar giant planets in multiple systems$({\it 1})$.
The masses of the surviving clumps, located between 3 and 20 AU, are in 
the range 2-6 Jupiter masses in
the adiabatic case and twice as big in the isothermal case.
Most of the 
clumps have orbits with final eccentricities  between 0.1 and 0.3 (Fig.2)
as many of the observed extrasolar planets (Fig.2).
The simulated protoplanets enter
a rather quiescent evolutionary phase (no more mergers occur) 
several orbital times before the end of the simulations.
However, inward migration might continue on 
time-scales longer than those explored here$({\it 10-12})$,
and orbits can also change due to interactions between the planets$(30)$.

This work shows that gravitational instability can 
actually form self-gravitating protoplanets and that long-lived
systems with masses and orbits consistent with those of extrasolar
planets arise. 
All this requires is to start with a marginally unstable disk 
($Q_{min}=1.4-1.5$) in which radiative cooling is efficient during the 
initial growth of the overdensities.
We tested that hotter disks starting from a considerably higher $Q_{min}$ 
($\sim 2$) can also become strongly unstable and form clumps if they are 
slowly  cooled to temperatures comparable to those used in the disks that 
start with $Q_{min} \sim 1.4({\it 31})$.
Therefore, clump formation does not depend on how the disk reaches the 
state used in our initial conditions. 
Ice giant planets, like Uranus and Neptune, might also be formed by the same 
mechanism after a strong ultraviolet flux from nearby bright stars 
has photoionized the envelopes of 
protoplanets more massive than Jupiter, leaving a mostly metallic core$({\it 20})$.
Because instabilities occur quickly,
future observations of planets around very young stars
will be a test for this model. In addition,
direct imaging of giant planets at large distances ($R >$ 50 AU) from the
stars$({\it 32})$ could also provide support to this model; 
as the outermost regions of the disk would be
even cooler (and surface density and angular velocity fall equally
with radius, i.e. as $r^{-3/2}$) $Q$ will be still decreasing at 
$R >$ 20 AU, and thus fragmentation should occur out to these large distances. 
On the contrary, Jupiter-like planets would not form at such large distances
in the core-accretion model because coagulation of planetesimals
into a solid core would take too long with the small surface densities
involved$({\it 33})$. Future observations of the gaseous medium in disks 
at different evolutionary stages, for example with the Space
Infrared Telescope Facility (SIRTF), will help constrain the
evolution of disk structure and will show whether it is consistent with
the gravitational instability picture.

\newpage

{\bf References and Notes}

1. G.W. Marcy, R. P. Butler, {\it Publ. Astr. Soc. Pac.}{\bf 112}, 137 (2000).

2. G.W. Marcy, R. P. Butler, {\it Annu. Rev. Astron. Astrophys} {\bf  36}, 57 (1998).

3. J.B. Pollack {\it et al.},{\it Icarus} {\bf 124}, 62 (1996).

4. J. Lissauer, {\it Nature} {\bf 409}, 23 (2001).

5. J. Bally., L. Testi, A. Sargent, J. Carlstrom, {\it Astron. J.} {\bf 116},   854 (1998).

6. C. Briceno {\it et al.}, {\it Science} {\bf 291}, 93 (2001).

7. K.E. Haisch, Jr., E.A. Lada, C.J. Lada, {\it Astrophys. J.} {\bf 553}, L153 (2001)

8. H.B. Throop, J. Bally, J.L.W. Esposito, M.J McCaughrean, {\it Science} {\bf 292}, 1686 (2001)

9. D.N.C. Lin, J.C.B. Papaloizou, {\it Astrophys. J.} {\bf 309}, 846 (1986)

10. G. Bryden, X. Chen, D.N.C. Lin, R.P. Nelson, J.C.B. Papaloizou, {\it Astrophys. J.} {\bf 514}, 344 (1999).

11. R.P. Nelson, J.C.B. Papaloizou, F. Masset, W. Kley, {\it Mon. Not. Rev. Astron. Soc.} {\bf 318}, 18 (2000).

12. H. Tanaka, T. Takeuchi, W. Ward, {\it Astrophys.J.} {\bf 565}, 1257 (2002).

13. A. P. Boss, {\it Astrophys. J.} {\bf 503}, 923 (1998).

14. S.J. Weidenschilling, {\it Astrophys. Space. Sci} {\bf 51}, 153, (1977).


15. A.P. Boss, {\it Science} {\bf 276}, 1836 (1997).

16. A.P. Boss, {\it Astrophys. J} {\bf 563}, 367 (2001).

17. A.P. Boss,{\it Ann. Rev,. Astron. Astrophys.} {\bf 26}, 53 (1998).

18. P. D'Alessio, N. Calvet, L. Hartmann, {\it Astrophys. J.} {\bf 553}, 321 (2001).

19. T. Guillot, {\it Science} {\bf 286}, 72 (1999).

20. A.P. Boss, G.W. Wetherill, N. Haghighipour, {\it Icarus} {\bf 156}, 291 (2002).

21. R.A. Gingold, J.J. Monaghan, {\it Mon. Not. R. Astron. Soc.} {\bf 181},
357 (1977). 

22. G. Laughlin, P. Bodenheimer, {\it Astrophys. J.} {\bf 436}, 335 (1994).

23. A.F. Nelson, W. Benz, A.F. Adams, D. Arnett, {\it Astrophys. J} {\bf 502}, 342 (1998).

24. A.F. Nelson, W. Benz, T.V. Ruzmaikina, {\it Astrophys. J.} {\bf 529}, 457 (2000).

25. B.K. Pickett, P. Cassen, R.H. Durisen, R. Link, {Astrophys. J.} {\bf 529}, 1034 (2000).

26. B.K. Pickett, R.H. Durisen, P. Cassen, C. Mejia, {Astrophys. J.} {\bf 540}, L95 (2000).

27. Artificial viscosity is used in both the momentum and energy equation
but its contribution is minimized when the local flow is purely shearing
by means of a correction term originally introduced by D.S. Balsara ({\it
Comp. Phys., {\bf 121}, 357 (1995))}; the latter term is necessary
to prevent SPH from generating a large artificial viscosity in a purely
shearing flow.
In addition, when the disk evolves following a locally isothermal 
equation of state any entropy generated by the viscosity term is assumed to be 
instantly radiated away.

28. The Jeans mass is always resolved by more than 
several thousand particles at the locations of the clumps; this resolution
is more than an order of magnitude higher than required to avoid artificial
fragmentation based on the criteria developed by M. Bate, A. Burkert ({\it Mon. Not. R. Astron. Soc.} {\bf 288}, 1060 (1997)) for three-dimensional SPH
simulations.

29. In this regime reversible heating arising from compressions is not
radiated away and the artificial viscosity creates irreversible 
heating in regions where shock waves occur. 

30. C. Terquem, J.C.B. Papaloizou, {\it Mon. Not. R. Astron. Soc.} {\bf 332}, L39 (2002).


31. For this test we increased the temperature of the most massive disk model 
(resolved by 200.000 particles) by a factor of 2, reaching 100 K in the outer region; 350 years after the start of the
simulations the disk was still fairly axisymmetric and we started cooling it
at the constant rate of about 0.2 Kelvin/year. The disk underwent increasingly
strong spiral instabilities as it was cooled down; these produced an inward
flux of mass and outward flux of angular momentum but overall the outer surface
density did not change significantly. Therefore, reaching a temperature only
slightly smaller than that used in the standard initial conditions (42 K as
opposed to 50 K) was enough to compensate for the decrease of surface density
and give rise to clump formation after about 650 years (cooling was stopped
when fragmentation was approached).

32. K.L. Luhman, R. Jayawardhana, {\it Astrophys. J.} {\bf 566}, 1132 (2002).

33. E.W. Thommes, M.J. Duncan, H.F.  Levison, {\it Astron. J.} {\bf 123}, 2862 (2002).


34.J.Stadel, J.Wadsley \&  D.Richardson, in "High Performance Computing Systems
and Applications", eds. M.J.Dimoupulos \& K.F.Lie, Kluwer Academic Publishers,
Boston, 501 (2002)

35. The authors thank Gregory Laughlin, David Hollenbach and Alan Boss for 
useful and stimulating discussions.
Simulations were carried out at the Pittsburgh Supercomputing Center and
at CINECA. This research was supported by a grant from the NSF and by
the NASA Astrobiology Institute.

\newpage

\medskip
\center
\epsfxsize=14truecm
\epsfbox{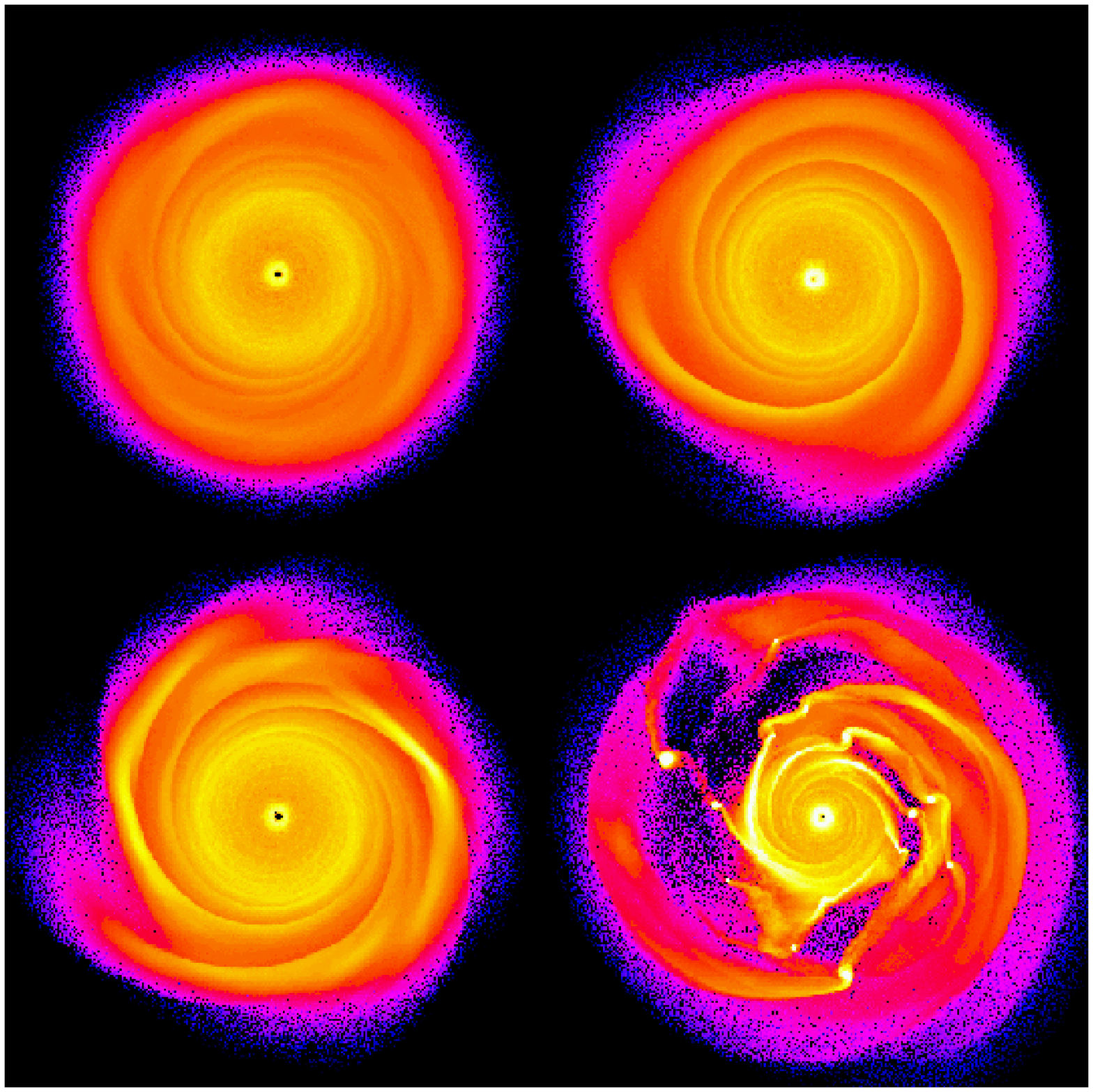}
\medskip
\figcaption[fig1.ps]{\label{fig:asymptotic}
\small{Snapshots of the simulations showing the protoplanetary disks 
seen face-on at different times. The colour-coded density on a logarithmic 
scale is shown out to 20 AU. Brighter colors trace higher densities;
the density ranges 
between $10^{-14}$ and $10^{-6}$ g/cm$^3$ is shown using
a logarithmic scale. 
The evolution of two disks having a different initial minimum
Toomre Q parameter, $Q_{min}$, are shown. The Toomre parameter at
a given disk location is defined as
$Q=\Omega v_s / \pi G\Sigma$, where $\Sigma$ is
the gas surface density, $\Omega$ is the angular velocity, $G$ is the
gravitational constant, and 
$v_s$ is the sound speed, $v_s=\sqrt{P/\rho}$, where $P$ is the pressure and $\rho$ is the 
density of the gas.  Disks have a surface density profile $\Sigma \sim r^{-3/2}$$({\it 12})$ and temperature profiles as in Boss$({\it 9-11})$.
Q reaches its minimum at $R > 10$ AU, where the temperature is 
as low as 50 K, while $Q> 4$ close to the inner disk boundary, where
the temperature is around $650$ K. 
The two upper panels show the disk 
with initial $Q_{min} \sim 1.75$ ($M_{disk}= 0.08 M_{\odot}$)
at $T=160$ yr (left) and $T=350$ yr (right)
, while the two lower panels show the disk with $Q_{min} \sim 1.4$ 
($M_{disk}=0.1 M_{\odot}$)
at $T=160$ yr and  $T=350$ yr. 
Gravity is softened on scales of 0.06 AU for disk particles.
The central stellar potential is exactly keplerian at 2.5 AU and it is
softened on smaller scales to speed up the computation. Both the central star
and the inner disk boundary are free to move. The simulations were 
performed with GASOLINE, a parallel N-body/SPH code in which gravity is computed using a binary tree$({\it 34})$.}}

\medskip

\newpage

\center
\epsfxsize=14truecm 
\medskip
\epsfbox{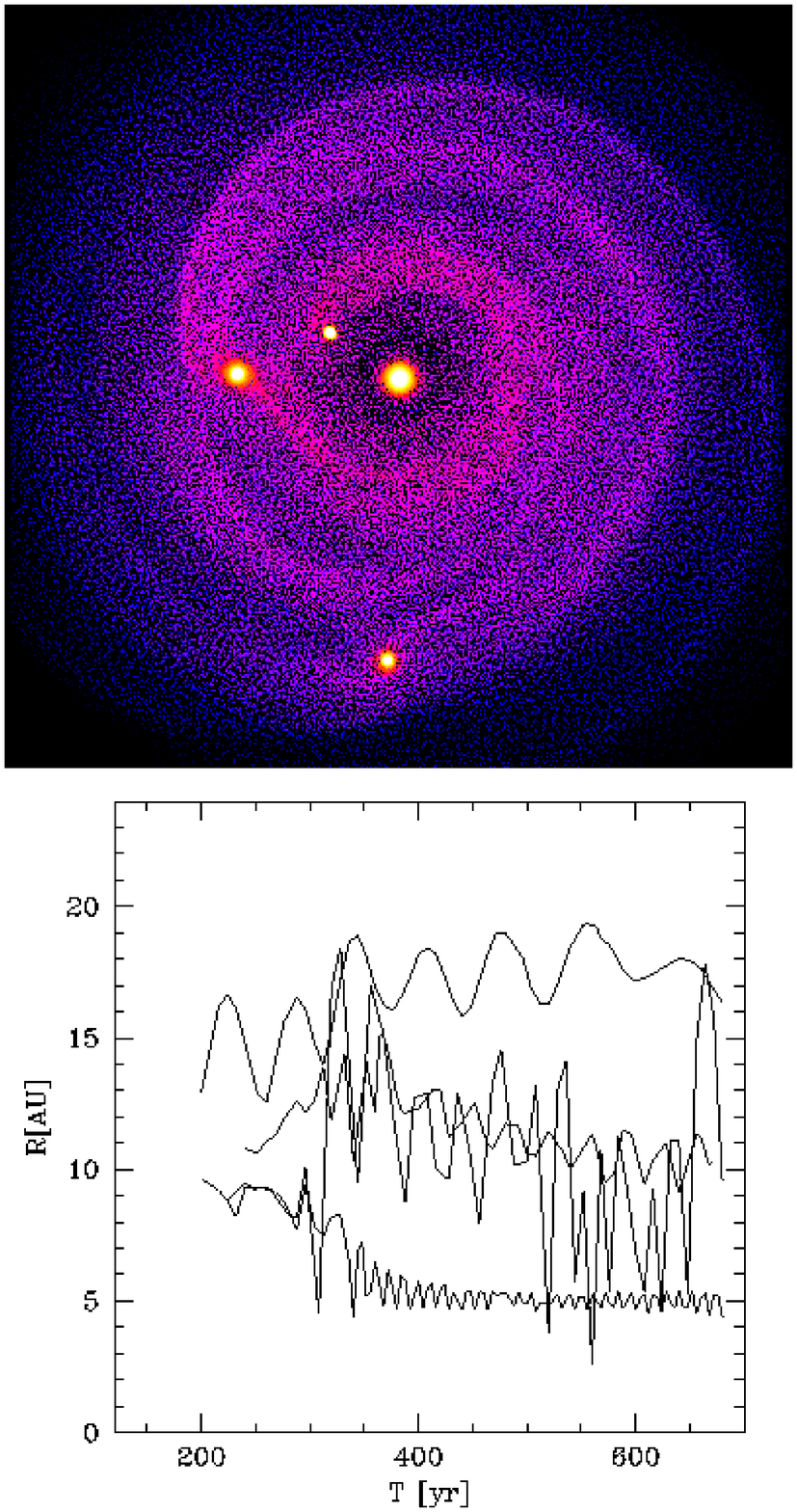}
\figcaption[neworbclumps.ps]{\label{fig:asymptotic2}
\small{Late stage of disk evolution. On top a face-on view of the
disk after 800 years is shown for the 200.000 particles simulation
in which the equation of state is switched to adiabatic after
about $200$ yr. The colour-coded logarithmic density (see Fig.1)
is shown out to 25 AU.
Three giant protoplanets are left and their orbital evolution is
shown in the plot at the bottom (thick lines) together with 
that of a clump from the simulation
where the equation of state is kept isothermal (thin line).
Each of the remaining clumps is the end result of a series of mergers;
the orbital evolution of the most massive progenitor is shown.  
The complex combination of torquing by the non-axisymmetric disk and 
interactions
with other clumps changes the orbital eccentricity and mean radius of the orbits. Overall the orbital evolution is considerably more complex
than the nearly-steady inward migration expected in light, 
axisymmetric disks,
eventually halting once the planet has cleared a gap$({\it 9-12})$.}}
\medskip

\end{document}